# A secure email login system using virtual password

Bhavin Tanti {bhavintanti@gmail.com}

Nishant Doshi {doshinikki2004@gmail.com}

*Abstract*—In today's world password compromise by some adversaries is common for different purpose. In ICC 2008 Lei et al. proposed a new user authentication system based on the virtual password system. In virtual password system they have used linear randomized function to be secure against identity theft attacks, phishing attacks, keylogging attack and shoulder surfing system. In ICC 2010 Li's given a security attack on the Lei's work. This paper gives modification on Lei's work to prevent the Li's attack with reducing the server overhead. This paper also discussed the problems with current password recovery system and gives the better approach.

*Index Terms*— Cryptography, Email attack, Security, Virtual password.

## I. INTRODUCTION

In the client server relate security system environment one of the most defensive e component is the client or user authentication module which allows the server to grant access and deny to others [1]. In today's there are many methods available like PIN, secret question, biometrics etc. out of all previously methods the PIN methods used widely due to less complex, less costly etc.

There are certain problem can happen with PIN problem. One of well known is there are static password so they can be stolen means stealing the client identity. Some attacks including phishing [2], malware (record the keystrokes) based attacks [3] and shoulder surfing attacks [4]. All of these attacks are previously described in [5].

The focus of this paper is the virtual password system which was proposed by Lei et al in [7], [8], secret little functions in [9] and by Li in [6]. The proposed virtual system claimed secure again all the attacks previously given but in [6] proved that with on an average by decoding 2 encrypted messages the virtual password system can be compromised. And the password can be useful to impersonate the user.

In email password recovery we use secondary email address to reset the password which was compromised. But the attacker can change the secondary email id after compromising the password. So this paper gives proposed change in this problem.

The rest of the paper is organized as follows. The next section gives background study or literature study which proposed in [7], [8]. In section III we discussed the how modified system can prevent the attack discussed in [6]. In section IV the system and usability with clients were explained. Related work and future expansion are given in section VI.

## II. BACKGROUND STUDY

### A. The concept of Virtual Password

The idea behind virtual password is to hiding the password by generating fresh password every time or random password every time. The server and user share a virtual password which was composed of two parts. 1) A fixed secret password $X=x_1,x_2,…x_n$, where each $x_i \in Z$ and Z will be set of all password characters. 2) For each login section server will generate or user will provide with random salt $Y= y_1, y_2, …y_n$, where $y_i \in Z$ and based on this user will enter a virtual password $K= k_1, k_2,…k_n=B(X, Y)$ to clear the authentication the process. So this protocol is common challenge-response method following based on secret key. In section III of [7] and section 3 of [8] Lei considered secret key is the fixed part so in [6] the authors had given a type of attack that possible to impersonate the user.

### B. The virtual password system

The randomized linear function will follow the steps as given below. The fixed password is given by $X=x_1,x_2,…x_n$ and a secret integer $a \in Z$. the integer chosen in such a way that $gcd(a,Z)=1$. We assume the $Z=\{0,1,…,Z-1\}$ i.e. the cardinality of set Z.

1. The server generates a random salt $Y = y_1 \cdots y_n$ and sends it to the user, where $y_i \in Z$.

2. The user generates a random integer $c \in Z$,

   calculates $K = k_1 \cdots k_n$ as follows:
   – $k_1 = B_1(x_1, a, y_1, c) = (ax_1 + y_1 + x_2 + c) \mod Z$;
   – $k_i = B_i(x_i, a, k_{i-1}, y_i, c) = (ak_{i-1} + y_i + x_i + c + x_{i1}) \mod Z$ for $2 \leq i \leq n$, where $I + 1 = ((i+1) \mod n) + 1$. Then, the user sends $K$ to the server.
3. For $c = 0, ..., Z-1$, the server calculates $K$ in the same way as in Step 2, and checks if it matches the response received from the user. If no any value of $c$ produces a match, reject the user; otherwise accept him/her.

Lei et al. claimed that using radome integer c, the virtual password system is secure against multiple observer login. In [6] authors that the above statement wrong and the secret fixed password can be compromised successfully. They show using example.

Here in all previous work there were some assumptions required to be made. If we assume that the random salt is provide at login time than at that time user had to calculate the K and this can be detected in phishing or shoulder surfing attack easily. If we assume that user will come with random salt and password K for that and at login time he/she enter both random salt and K and server will verify by decrypting K and compare with random salt provided by user , if both match than user successfully login. The other assumption is server will not records the previous random generated salts so reply attacks can be possible.

### III. PROPOSED WORK

#### A. Modified Virtual password system

There are several ways we can defend the attack given by [6]. If we can send the value of c with each transaction than attacker does not know that for which particular c value the particular K value associate. So if we provide wrong c value for K value than server will know that attacker is trying to gain the access so it will deny the login.

Here server will check for message that for which value of c , $c \in Z$ the K is built. So if we assume the sufficiently large value of Z than the processing time of server will be increase in the distributed environment where lots of users are connected to server. So in that case sending c value with K in the encrypted form will save the server time. Another advantage is server will record all previous c values used between user for the same password and after login user can see this values with date and time so reply attack using same c value not possible.

Now if user generates the random number every time than we do not require c value so modified algorithm is as follow.

The randomized linear function will follow the steps as given below. The fixed password is given by $X = x_1, x_2, ... x_n$ and a secret integer $a \in Z$. the integer chosen in such a way that $gcd(a, Z) = 1$. We assume the $Z = \{0, 1, ..., Z-1\}$ i.e. the cardinality of set Z.

1 The user generates a random salt $Y = y_1 ... y_n$

   and encrypts using public key of server , where $y_i \in Z$.

2 The user calculates $K = k_1 ... k_n$ as follows:
   – $k_1 = B_1(x_1, a, y_1) = (ax_1 + y_1 + x_2 + c) \mod Z$;
   – $k_i = B_i(x_i, a, k_{i-1}, y_i) = (ak_{i-1} + y_i + x_i + x_{i+1}) \mod Z$ for $2 \leq i \leq n$, where $i + 1 = ((i+1) \mod n) + 1$. Then, the user sends K and encrypted random number to the server.
3 The server decrypts the random number and calculates K in the same way as in Step 2, and checks if it matches the response received from the user. If no match, reject the user; otherwise accept him/her.

In above steps we are not using random number c. the other way is to add one more step in previously algorithm, send value of c with current time stamp decrypt under public key of server. Server will cross verify the value of c so no need to check all values between 0 to Z-1.

#### B. Email password system

The problem with current email password reset/recovery system is given in introduction. If we want more security as reasonable cost than we require one more password that will be used to protect the secondary email id. So if the password for email id is compromised than attacker cannot change the secondary email id without as he not has the second password. So in other words we can say that second password only come in picture when user want to change the secondary email id. Suppose that user forgot the second password than to recover user simply make request and password or link to reset password will be send to secondary email id.

Now consider the other scenario, attacker had broken or get the password for some email so now he had access to all that email system for which the hacked email address were used for secondary email id. This attack cannot prevent by the above modified system.

### IV. SYSTEM AND USABILITY

#### A. Modified Virtual password system

In the [7] and [8] the concept of system and usability were discussed. The main aim of this is how quickly user can adapt the system. If we assume the user had mobile phone or laptop or some palmtop devices than user can simple installed the application which is freely available on internet and then enter secret key and random number will be generated by application itself and then application give final key K which user enter at login time. So if we assume that user do not have any of above devices than it's depending on user the ability to do mentally. So the modified system had same problem as previous system.

*B. Email password system*

In the present email system user has to give one password for doing all the stuff. But in new system user may require to remember both password. We assume that user will enter/select both the password at registration time.

V. CONCLUSION AND FUTURE WORK

This paper try to modify the existing scheme to prevent attack and give the minor change in email password reset system in order to get better security. In future may be another attack can be possible or we can minimize the overall length as well as can have better scheme for email password reset so the attack which possible in proposed system cannot possible in future system.